\documentclass[conference]{IEEEtran}
\IEEEoverridecommandlockouts

\ifCLASSINFOpdf
\else
   \usepackage[dvips]{graphicx}
\fi
\usepackage{url}
\usepackage{graphicx}

\usepackage{textcomp}
\usepackage{stfloats}
\usepackage{verbatim}
\usepackage{cite}
\usepackage{amsmath,amssymb,amsfonts}
\usepackage{algorithm}
\usepackage{algorithmic}  
\usepackage{graphicx}
\usepackage{float}
\usepackage{textcomp}
\usepackage{xcolor}
\usepackage{bbm}
\usepackage{array}
\usepackage{rotating}
\usepackage{gensymb}
\usepackage{comment}
\usepackage[acronym]{glossaries}
\usepackage{multirow}
\usepackage{lscape}
\usepackage{makecell}
\usepackage[inline]{enumitem}
\usepackage{svg}
\usepackage{hyperref}
\hypersetup{hidelinks}
\usepackage[font=small,labelsep=period,justification=centering]{caption}
\usepackage{subcaption}
\usepackage{amsmath}
\usepackage{amsthm}
\usepackage[noabbrev, capitalize]{cleveref}
\usepackage[normalem]{ulem}
\usepackage{tabularx}
\usepackage{array}

\DeclareMathOperator*{\argmin}{arg\,min}

\usepackage{etoolbox} 
\makeatletter
\patchcmd{\subsubsection}{\itshape}{\bfseries}{}{}
\makeatother

\newtheoremstyle{mystyle}%
  {1pt}    
  {1pt}    
  {\itshape} 
  {}        
  {\bfseries} 
  {.}       
  {3pt}     
  {}        

\theoremstyle{plain}

\crefname{proposition}{Proposition}{Propositions}
\theoremstyle{plain}

\crefname{definition}{Definition}{Definitions}

\newcommand{\nocontentsline}[3]{}
\newcommand{\tocless}[2]{\bgroup\let\addcontentsline=\nocontentsline#1{#2}\egroup}

\def\BibTeX{{\rm B\kern-.05em{\sc i\kern-.025em b}\kern-.08em
    T\kern-.1667em\lower.7ex\hbox{E}\kern-.125emX}}

\allowdisplaybreaks

\begin{document}

\title{Doppler-Only Extended Target Tracking in a Distributed Antenna System}

\author{\IEEEauthorblockN{Baptiste Sambon, Gilles Monnoyer, Luc Vandendorpe, and Claude Oestges \thanks{Baptiste Sambon is a Research Fellow of the Fonds de la Recherche Scientifique - FNRS.}}\\
\IEEEauthorblockA{ICTEAM, UCLouvain - Louvain-la-Neuve, Belgium\\}
email: firstname.lastname@uclouvain.be
}

\maketitle

\begin{abstract}

In modern radar and Integrated Sensing and Communication (ISAC) systems, the conventional point-scatterer assumption becomes increasingly inaccurate due to higher spatial resolution. In this context, Extended Target Tracking (ETT) has grown interest in recent years. However, most existing ETT works are designed for monostatic systems due to the fine time synchronisation or per-node antenna arrays required for ranging and angle estimation in multistatic scenarios. 
In contrast, this paper proposes a novel approach to extended target tracking in a distributed antenna system, where only bistatic Doppler frequency measurements are available. 
Doppler-only tracking is particularly attractive as it avoids the two aforementioned issues.
To address this Doppler-only extended target tracking problem, we develop a modified Rao-Blackwellised Particle Filter (RBPF) that replaces the standard Kalman update with a particle-specific Extended Kalman Filter update followed by a relinearisation step, handling the non-linearities of both the rigid extended target dynamics and the observation model.
Numerical simulations show the effectiveness of the proposed algorithm for accurate extended target tracking under this Doppler-only setting.

\end{abstract}


\IEEEpeerreviewmaketitle


\section{Introduction}
\label{sec:introduction}

Target tracking is a longstanding problem in radar and, more recently, in Integrated Sensing and Communication (ISAC) systems, aiming to jointly perform radar sensing and communication \cite{liu_integrated_2022}, with applications ranging from automotive to maritime and aerial scenarios \cite{han_improving_2025, fowdur_elliptical_2021, song_trajectory_2026}. As the spatial resolution of modern sensing systems increases, the conventional point-target assumption becomes less accurate, motivating growing interest in Extended Target Tracking (ETT), where targets are modeled as spatially extended objects \cite{thormann_extended_2018, steuernagel_extended_2025}.

In automotive ETT, Doppler measurements are often combined with range and/or angle information to improve tracking, e.g., using Gaussian process~\cite{thormann_extended_2018}, direct scattering~\cite{knill_direct_2016}, or B-spline models~\cite{han_improving_2025}. However, these approaches typically assume monostatic (co-located) sensing, whereas distributed (multistatic) systems require tight inter-node synchronisation for ranging or per-node antenna arrays for angle estimation.



On the other hand, distributed Doppler-only tracking, i.e. tracking using Doppler frequency alone without range or angle measurements, is particularly attractive because it does not require fine time synchronisation between distributed nodes \cite{krysik_doppler-only_2014, maasdorp_target_2015, peled-eitan_missile_2024}.
Besides, Doppler frequency can often be estimated more accurately than range with waveforms poorly suited to range estimation.
This is typically the case in passive radar systems exploiting existing communication or broadcast signals (e.g., GSM or FM) \cite{krysik_doppler-only_2014, maasdorp_target_2015} but also in active systems relying on narrowband or unmodulated waveforms, such as continuous-wave radars \cite{richter_tracking_2022} or ISAC systems reusing communication waveforms for sensing.
The deployment of such systems is facilitated for those reasons, and field trials have also shown the feasibility of Doppler-only tracking with low-cost distributed receivers~\cite{maasdorp_target_2015}. 
However, existing Doppler-only tracking approaches predominantly rely on the point-target assumption, associating the measured Doppler with the translation velocity of a single scattering point.

To the best of the authors' knowledge, the problem of tracking an extended target using solely bistatic Doppler frequency measurements in a distributed radar/ISAC system has not been studied in the literature, despite its practical relevance. This gap is addressed in this paper, where we adopt an elliptical target model, which is a common choice in the literature, as it is particularly relevant for automotive, aerial and maritime applications~\cite{fowdur_elliptical_2021, steuernagel_extended_2025}. 


A natural approach to solve this Doppler-only extended target tracking problem would be to directly apply standard non-linear filters such as Extended Kalman Filter (EKF) \cite{snyder_introduction_2014}, Unscented Kalman Filter (UKF) \cite{julier_unscented_2004}, or Particle Filter (PF) \cite{doucet_sequential_2013}. However, none of these methods is well suited to our problem. 
The EKF and UKF are known to perform poorly in highly non-linear settings due to the required linearisation, which is the case here due to the non-linearities in both the rigid-body motion of the target (process model) and the Doppler-only measurements (observation model). The PF, on the other hand, does not suffer from linearisation issues, but must explore a high-dimensional state space, which typically requires a large number of particles (and thus computational resources) to obtain an accurate posterior approximation. 

Among ETT approaches, Rao-Blackwellised Particle Filters (RBPFs) have been used to handle non-linear processes and observation models~\cite{steuernagel_extended_2025, knill_direct_2016}. Standard RBPFs partition the state into a non-linear component and a conditionally linear one, reducing the dimension explored by particles. Here, however, the bistatic Doppler observation remains non-linear in both components, preventing direct application of the standard RBPF.


Based on the aforementioned works, the contributions of this paper are summarised as follows. 
\begin{itemize}
   \item We formulate the problem of tracking a rigid elliptical extended target in a distributed radar/ISAC system using only bistatic Doppler frequency measurements.
   \item We develop a  modified Rao-Blackwellised Particle Filter (RBPF) algorithm adapted for our problem, in which, unlike in a standard RBPF, the Kalman filter update is replaced by a particle-specific Extended Kalman Filter (EKF) update, and a relinearisation step is performed after the update to improve the accuracy. This allows to handle the non-linearities arising from both the rigid-body target dynamics and the observation model. 
   \item We derive in closed form the Jacobian of the bistatic Doppler frequency with respect to the target state, which is used in the update step of the tracking algorithm.
   \item We validate the algorithm through simulations, showing it outperforms standard tracking algorithms in terms of Intersection over Union, with a low number of particles.
\end{itemize}


\section{System Model}
\label{sec:system_model}

\begin{figure}
\centering
\includegraphics[width=0.5\linewidth, trim=10 5 10 5, clip]{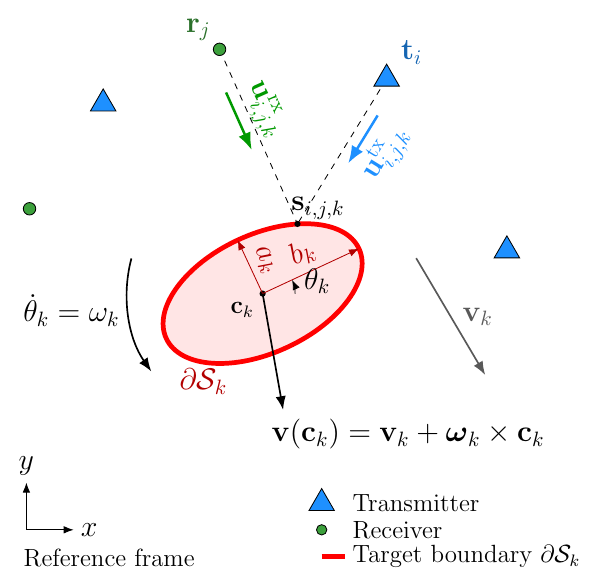}
\caption{Distributed radar/ISAC system illuminating a single extended target modeled as a rigid elliptical scatterer. }
\label{fig:system_model}
\end{figure}

As shown in Fig.~\ref{fig:system_model}, we consider a distributed radar/ISAC system with $N_T$ transmit antennas at known, fixed positions $\{ \mathbf{t}_i\}_{i=1}^{N_T} \subset \mathbb{R}^{2} $ and $N_R$ receive antennas at $\{ \mathbf{r}_j\}_{j=0}^{N_R-1} \subset \mathbb{R}^{2}$. A single extended target is modeled at each time step $k$ (sampling interval $\Delta t$) as a rigid elliptical scatterer with boundary 
\begin{equation}
\partial\mathcal{S}_k \triangleq \big\{ \mathbf{c}_k + \mathbf{R}(\theta_k)[b_k\cos\varphi,\, a_k\sin\varphi]^T : \varphi\in[0,2\pi) \big\},
\label{eq:elliptical_target_boundary}
\end{equation}
where $\mathbf{c}_k$ is the target centroid, $\mathbf{R}(\theta)=\big[\begin{smallmatrix}\cos\theta&-\sin\theta\\sin\theta&\cos\theta\end{smallmatrix}\big]$ is a rotation matrix, $\theta_k$ is the orientation of the target-fixed $b_k$-axis with respect to the $x$-axis of the global reference frame, and $a_k$ and $b_k$ are the semi-axis lengths.

The rigid-body motion is parameterised by a linear component $\mathbf{v}_k=(v_{x,k},v_{y,k})^T$ and an angular velocity $\omega_k$, as described in Section~\ref{sec:kinematic_model}. The target state vector at a given time step $k$ is thus defined as
\begin{equation}
\mathbf{x}_k \triangleq \big[\mathbf{c}_k^T, a_k, b_k, \mathbf{v}_k^T, \omega_k, \theta_k \big]^T \in \mathbb{R}^{8}.
\end{equation}

\subsection{Kinematic Model}
\label{sec:kinematic_model}

The velocity of any point $\mathbf{s}$ of the target, is conventionally expressed with respect to the target centroid $\mathbf{c}_k$ as
\begin{equation}
   \mathbf{v}(\mathbf{s}) = \mathbf{v}(\mathbf{c}_k) + \boldsymbol{\omega}_k\times(\mathbf{s}-\mathbf{c}_k),
\end{equation}
where \(\mathbf{v}(\mathbf{c}_k)\) is the centroid velocity and $\boldsymbol{\omega}_k = [0,0,\omega_k]^T$ is the angular velocity vector. The same velocity can equivalently be written as
\begin{equation}
   \mathbf{v}(\mathbf{s}) = \mathbf{v}_k + \boldsymbol{\omega}_k\times\mathbf{s},
\label{eq:rigid_body_velocity}
\end{equation}
where
\begin{equation}
\mathbf{v}_k = \mathbf{v}(\mathbf{c}_k) - \boldsymbol{\omega}_k\times\mathbf{c}_k
\end{equation}
is the linear component of the spatial rigid-body velocity \cite{kellner_tracking_2016, murray_mathematical_2017}. 
This equivalent parametrisation in terms of $(\mathbf{v}_k,\omega_k)$ is adopted in the state vector $\mathbf{x}_k$ to describe the target rigid-body motion. It is particularly convenient for tracking, since a nearly constant $(\mathbf{v}_k,\omega_k)$ model naturally encompasses trajectories typically encountered in vehicular applications, e.g., straight motion, smooth curves, and constant-radius turns.

\subsection{Specular Point Model}

Following the extended target model developed in \cite{sambon_electromagnetic_2025}, each transmit-receive antenna pair $(i,j)$ illuminates a single specular point $\mathbf{s}_{i,j,k} \in \partial\mathcal{S}_k$ on the target boundary at time step $k$. Recalling from~(1) that any point on $\partial\mathcal{S}_k$ can be written as $\mathbf{q}(\varphi) \triangleq \mathbf{c}_k + \mathbf{R}(\theta_k)[b_k\cos\varphi,\ a_k\sin\varphi]^T$ for some $\varphi \in [0,2\pi)$, the specular point can equivalently be written as $\mathbf{s}_{i,j,k} = \mathbf{q}(\varphi_{i,j,k})$, i.e. as a single scalar unknown $\varphi_{i,j,k}$ rather than a free point in $\mathbb{R}^2$. This point is defined as the one on the target boundary that satisfies the law of reflection, and can be computed as the boundary angle minimising the total bistatic path length according to Fermat's principle~\cite{miller_fermats_2009}, i.e. 
\begin{equation}
\varphi_{i,j,k} = \argmin_{\varphi\in[0,2\pi)} g\big(\mathbf{q}(\varphi) ; \mathbf{t}_i, \mathbf{r}_j\big) \text{ and } \mathbf{s}_{i,j,k} = \mathbf{q}(\varphi_{i,j,k}),
\label{eq:specular_point}
\end{equation}
where $g(\mathbf{s} ; \mathbf{t}_i, \mathbf{r}_j) = \big( \|\mathbf{s}-\mathbf{t}_i\| + \|\mathbf{s}-\mathbf{r}_j\| \big)$.
Equation~\eqref{eq:specular_point} can admit two solutions, corresponding to the two points on the target boundary that satisfy the law of reflection. The specular point is chosen as the one that fulfills the outgoing condition, i.e. the one that is "visible" from both the transmit and receive antennas.
Pairs $(i,j)$ for which no specular point exists at time $k$ (e.g., no visible specular point) are discarded from the corresponding observation set $\mathcal{P}_k$. 
The specular point can be computed numerically using a gradient descent algorithm. 

\subsection{Observation Model}

Evaluating the rigid-body velocity \eqref{eq:rigid_body_velocity} at the instantaneous specular location $\mathbf{s}_{i,j,k}$ gives
\begin{equation}
\mathbf{v}_{i,j,k} = \mathbf{v}_k + \boldsymbol{\omega}_k \times\mathbf{s}_{i,j,k}.
\label{eq:speed_specular_point}
\end{equation}
The bistatic Doppler frequency $f_{i,j,k}$ observed at the receive antenna $j$ for the transmit antenna $i$ is then given by
\begin{equation}
f_{i,j,k} = \frac{1}{\lambda} \mathbf{v}_{i,j,k}^T \big( \mathbf{u}_{i,j,k}^{\text{tx}} + \mathbf{u}_{i,j,k}^{\text{rx}} \big),
\label{eq:bistatic_doppler}
\end{equation}
where $\lambda $ is the wavelength, and $\mathbf{u}_{i,j,k}^{\text{tx}}$ and $\mathbf{u}_{i,j,k}^{\text{rx}}$ are the unit vectors pointing from the transmit and receive antennas, respectively, to the specular point. 
Note that the bistatic Doppler frequency is a nonlinear function of the target state vector $\mathbf{x}_k$ through the instantaneous specular point $\mathbf{s}_{i,j,k}$ at time $k$ appearing in both $\mathbf{v}_{i,j,k}$ and the unit vectors $\mathbf{u}_{i,j,k}$.

Stacking \eqref{eq:bistatic_doppler} for all antenna pairs $(i,j)\in\mathcal{P}_k$ yields the non-linear observation vector $\mathbf{z}_k \in \mathbb{R}^{|\mathcal{P}_k|}$ with 
\begin{equation}
\mathbf{z}_k = \mathbf{h}(\mathbf{x}_k) + \mathbf{n}_k, \qquad \mathbf{n}_k \sim \mathcal{N}(\mathbf{0},\sigma_n^2\mathbf{I}_{|\mathcal{P}_k|})
\label{eq:observation}
\end{equation}
where $\mathbf{h}(\cdot)$ is the non-linear observation function and $\mathbf{n}_k$ is the observation noise vector, assumed to be Gaussian with zero mean and covariance $\sigma_n^2\mathbf{I}_{|\mathcal{P}_k|}$.

In practice, those Doppler frequency measurements can for example be obtained from opportunistic signals in passive systems (e.g., FM, GSM signals \cite{krysik_doppler-only_2014,maasdorp_target_2015}) by comparing them to a reference system, by performing a Short-Time Fourier Transform (STFT) on the received signal (with the clutter at the null frequency), from subcarrier phase shifts in OFDM systems \cite{liu_integrated_2022}, or from the phase derivative of the received signal in Continuous-Wave (CW) systems \cite{richter_tracking_2022}.

\subsection{State-Space Model}

We assume that the target state vector $\mathbf{x}_k$ evolves with constant translation and angular velocities, and cannot be deformed (i.e., $a_k$ and $b_k$ are constant). The state-space model is thus given by
\begin{subequations}
\label{eq:process}
\begin{align}
\mathbf{c}_{k} 
&= \mathbf{c}_{k-1} + \Delta_t
\left(\mathbf{v}_{k-1} + \boldsymbol{\omega}_{k-1} \times \mathbf{c}_{k-1}\right)
+ \mathbf{w}_k^{\mathrm{c}}, \\
a_{k} 
&= a_{k-1} , 
\qquad
b_{k} = b_{k-1} , \\
\mathbf{v}_{k} 
&= \mathbf{v}_{k-1} + \mathbf{w}_k^{\mathrm{v}}, \\
\omega_{k} 
&= \omega_{k-1} + \mathrm{w}_k^{\omega},
\qquad
\theta_{k} = \theta_{k-1} + \Delta_t \omega_{k-1} + \mathrm{w}_k^{\theta}, 
\end{align}
\end{subequations}
compactly written as
\begin{equation}
\mathbf{x}_{k} = \mathbf{f}(\mathbf{x}_{k-1}) + \mathbf{w}_k, \qquad \mathbf{w}_k \sim \mathcal{N}(\mathbf{0},\mathbf{Q}),
\end{equation}
where $\mathbf{w}_k$ is the process noise vector, assumed to be Gaussian with zero mean and covariance $\mathbf{Q} = \text{diag}(\sigma_c^2\mathbf{I}_2, 0, 0, \sigma_v^2\mathbf{I}_2, \sigma_\omega^2, \sigma_\theta^2)$.

Equations \eqref{eq:observation} and \eqref{eq:process} thus define non-linear observation and state-space models for the extended target tracking problem, which can be solved using the tracking algorithm described in the next section (Section~\ref{sec:tracking_algorithm}).


\section{Proposed Tracking Algorithm}
\label{sec:tracking_algorithm}

To deal with the non-linearities in the observation and process models, we develop a modified Rao-Blackwellised Particle Filter (RBPF) to estimate the target state vector $\mathbf{x}_k$ at each time step $k$. 
First, we briefly introduce the standard RBPF in \cref{sec:rbpf_background} and why it must be modified in our case, then we describe the proposed algorithm in more detail. 

\subsection{Background on Rao-Blackwellised Particle Filtering and State Partition}
\label{sec:rbpf_background}

Consider a generic nonlinear model with a state vector $\tilde{\mathbf{x}}_k$ that can be partitioned in two components, $\tilde{\mathbf{x}}_k = [\tilde{\boldsymbol{\eta}}_k^\top, \tilde{\boldsymbol{\xi}}_k^\top]^\top$. The RBPF assumes that both the state-space and observation models are linear in $\tilde{\boldsymbol{\xi}}_k$ given $\tilde{\boldsymbol{\eta}}_k$,
i.e. 
\begin{equation}
\begin{aligned}
\tilde{\boldsymbol{\xi}}_k
&=
\tilde{\mathbf{F}}(\tilde{\boldsymbol{\eta}}_{k-1})
\tilde{\boldsymbol{\xi}}_{k-1}
+
\tilde{\mathbf{f}}(\tilde{\boldsymbol{\eta}}_{k-1})
+
\tilde{\mathbf{w}}_k^{\tilde{\boldsymbol{\xi}}},
\\
\tilde{\mathbf{z}}_k
&=
\tilde{\mathbf{C}}(\tilde{\boldsymbol{\eta}}_k)
\tilde{\boldsymbol{\xi}}_k
+
\tilde{\mathbf{h}}(\tilde{\boldsymbol{\eta}}_k)
+
\tilde{\mathbf{n}}_k , 
\end{aligned}
\end{equation}
where $\tilde{\mathbf{F}}(\cdot)$ and $\tilde{\mathbf{C}}(\cdot)$ are matrix-valued functions of $\tilde{\boldsymbol{\eta}}$,
$\tilde{\mathbf{f}}(\cdot)$ and $\tilde{\mathbf{h}}(\cdot)$ are arbitrary functions of $\tilde{\boldsymbol{\eta}}$, 
$\tilde{\mathbf{z}}_k$ is the observation vector,
and $\tilde{\mathbf{w}}_k^{\tilde{\boldsymbol{\xi}}}$ and $\tilde{\mathbf{n}}_k$ are Gaussian noise vectors. 

The posterior distribution of the state vector $\tilde{\mathbf{x}}_k$ given the observations $\tilde{\mathbf{z}}_{1:k}$ can be factorised as
\begin{equation}
p(\tilde{\mathbf{x}}_k | \tilde{\mathbf{z}}_{1:k}) = p(\tilde{\boldsymbol{\xi}}_k, \tilde{\boldsymbol{\eta}}_k | \tilde{\mathbf{z}}_{1:k})
= p(\tilde{\boldsymbol{\eta}}_k | \tilde{\mathbf{z}}_{1:k}) p(\tilde{\boldsymbol{\xi}}_k | \tilde{\boldsymbol{\eta}}_k, \tilde{\mathbf{z}}_{1:k}).
\end{equation}
In this case, the $p(\tilde{\boldsymbol{\eta}}_k | \tilde{\mathbf{z}}_{1:k})$ distribution can be approximated by a Particle Filter (PF) \cite{doucet_sequential_2013}, each particle $m$ carrying a candidate state $\tilde{\boldsymbol{\eta}}_k^{(m)}$. 
Since both the state-space and observation models are linear in $\tilde{\boldsymbol{\xi}}_k$ conditioned on $\tilde{\boldsymbol{\eta}}_k$, the corresponding conditional distribution $p(\tilde{\boldsymbol{\xi}}_k | \tilde{\boldsymbol{\eta}}_{1:k}^{(m)}, \tilde{\mathbf{z}}_{1:k})$ can be characterised analytically, for each particle $\tilde{\boldsymbol{\eta}}_k^{(m)}$, by its conditional estimate $\tilde{\boldsymbol{\xi}}_k^{(m)}$ and associated covariance, using the classical Kalman filter equations. 
These conditional estimates are then combined through the particle weights to form the overall state estimate. 
This approach is known as Rao-Blackwellised Particle Filtering \cite{doucet_rao-blackwellised_2000}, and offers the following advantages. 
\begin{itemize}
   \item Since $\tilde{\boldsymbol{\xi}}_k$ is computed analytically rather than sampled, the variance of the estimate is reduced (by the Rao-Blackwell theorem), and fewer particles are required to achieve a given accuracy compared to a standard particle filter over the full state vector $\tilde{\mathbf{x}}_k$.
   \item The Kalman filter update for $\tilde{\boldsymbol{\xi}}_k$ is computationally efficient, and MMSE-optimal given the sampled $\tilde{\boldsymbol{\eta}}_k^{(m)}$.
\end{itemize}

These benefits rely on the structure of the state and observations models, which must be both linear in $\tilde{\boldsymbol{\xi}}_k$ given $\tilde{\boldsymbol{\eta}}_k$. 

In our case, we can partition the target state vector $\mathbf{x}_k$ into 
\begin{equation}
\boldsymbol{\xi}_k \triangleq \big[\mathbf{c}_k^T, a_k, b_k, \mathbf{v}_k^T \big]^T \in \mathbb{R}^{6}, \qquad
\boldsymbol{\eta}_k \triangleq \big[\omega_k, \theta_k\big]^T \in \mathbb{R}^{2},
\end{equation}
since the state-space model \eqref{eq:process} is linear in $\boldsymbol{\xi}_k$ given $\boldsymbol{\eta}_k$. 
However, the observation model remains non-linear in $\boldsymbol{\xi}_k$ given $\boldsymbol{\eta}_k$, since \eqref{eq:bistatic_doppler} is a non-linear function of the specular point $\mathbf{s}_{i,j,k}$ and that the latter depends on both $\boldsymbol{\xi}_k$ and $\boldsymbol{\eta}_k$ through the boundary \eqref{eq:elliptical_target_boundary} and the specular point equation \eqref{eq:specular_point}, in a non-linear manner.

This has three main consequences that differ and extend the standard RBPF. First, the Kalman filter update for $\boldsymbol{\xi}_k$ must be replaced by an Extended Kalman Filter (EKF) update \cite{snyder_introduction_2014}, which linearises the observation model around an estimate of $\boldsymbol{\xi}_k$ to compute the Kalman gain. 
Second, in a standard RBPF, the observation matrix $\tilde{\mathbf{C}}(\tilde{\boldsymbol{\eta}}_k)$ depends on $\tilde{\boldsymbol{\eta}}_k$ only, so it can be computed directly from the predicted $\tilde{\boldsymbol{\eta}}_k^{(m)}$ of each particle, independently of $\tilde{\boldsymbol{\xi}}_k^{(m)}$. In our case, however, the linearised observation model also depends on $\boldsymbol{\xi}_k$, so its Jacobian must be evaluated using both the predicted $\boldsymbol{\xi}_{k|k-1}^{(m)}$ and $\boldsymbol{\eta}_{k|k-1}^{(m)}$ of each particle, making the EKF update particle-specific. 

Third, because the specular points $\mathbf{s}_{i,j,k}$ (and hence the observation model) depend on the state at which they are evaluated, the first EKF update may rely on a linearisation point that is not sufficiently close to the posterior state. 
To mitigate the resulting linearisation error, after the EKF update, we recompute the specular points, the observation function, and its Jacobian at the updated conditional state $\boldsymbol{\xi}_k^{(m)}$ for each particle $m$. This post-update relinearisation is used for the subsequent particle-weight computation. This procedure is inspired by the iterated EKF \cite{liang-qun_iterated_2005}, and allows the observation model to be relinearised around a state closer to the posterior estimate. Its impact on the tracking accuracy is evaluated in Section~\ref{sec:tracking_algorithm_comparison}.


The modified RBPF operations (prediction, update, estimate, and resampling) at a given time step $k$ are described in \cref{alg:rbpf_step} and detailed in the following subsections. 

\begin{algorithm}[t]
\caption{Modified RBPF recursion at time step $k$}
\label{alg:rbpf_recursion}
\begin{algorithmic}[1]

\FOR{$m=1, \dots,  N_P$}

   \STATE Predict $\boldsymbol{\eta}_k^{(m)}$ using \eqref{eq:pred_eta}
   \STATE Predict  $\boldsymbol{\xi}_{k|k-1}^{(m)}$ and $\mathbf{P}_k^{(m)}$ using \eqref{eq:pred_xi}-\eqref{eq:pred_P}
   \STATE Solve \eqref{eq:specular_point} for all $(i,j)\in\mathcal{P}_k$ to get $\{ \mathbf{s}_{i,j,k} \}$, and $\mathbf{h}(\cdot)$
   \STATE Compute $\mathbf{H}_k^{(m)}$ using \eqref{eq:H_v}-\eqref{eq:grad_f}
   \STATE Update $\boldsymbol{\xi}_k^{(m)}$ and $\mathbf{P}_k^{(m)}$ using \eqref{eq:A}-\eqref{eq:Pupd}
   \STATE Relinearise: recompute $\mathbf{h}(\cdot)$ and $\mathbf{H}_k^{(m)}$ at the updated state ${\boldsymbol{\xi}}_k^{(m)}$ using \eqref{eq:specular_point} and \eqref{eq:H_v}-\eqref{eq:grad_f}
   \STATE Update the particle weight $w_k^{(m)}$ using \eqref{eq:weight}

\ENDFOR

\STATE Normalise the weights $w_k^{(m)} \leftarrow w_k^{(m)}/\sum_{m=1}^{N_P} w_k^{(m)}$
\STATE Compute the target state estimate $\hat{\mathbf{x}}_k$ via \eqref{eq:state_estimate}

\IF{$N_\text{eff} < N_\text{th}$}
   \STATE Resample the particles using systematic resampling \cite{kitagawa_monte_1996}
   \STATE \vspace{-\baselineskip} Reset the weights $w_k^{(m)} = 1/N_P$ for all particles
\ENDIF

\end{algorithmic}
\label{alg:rbpf_step}
\end{algorithm}

\subsection{Prediction Step}

The distribution of the target state vector $\mathbf{x}_k$ is thus represented by $N_P$ weighted particles $\{\boldsymbol{\eta}_k^{(m)},{\boldsymbol{\xi}}_k^{(m)},\mathbf{P}_k^{(m)},w_k^{(m)}\}_{m=1}^{N_p}$: each particle carries a weight $w_k^{(m)}$ and a Gaussian distribution $\mathcal{N}(\boldsymbol{\xi}_k^{(m)},\mathbf{P}_k^{(m)})$ for the $\boldsymbol{\xi}_k^{(m)}$ component. 
For each particle $m$, 
\begin{align}
\omega_k^{(m)} &= \omega_{k-1}^{(m)} + \mathrm{r}_k^{\omega}, \qquad
\theta_k^{(m)} = \theta_{k-1}^{(m)} + \Delta t\,\omega_{k-1}^{(m)} + \mathrm{r}_k^{\theta}, \label{eq:pred_eta} \\
\boldsymbol{\xi}_{k|k-1}^{(m)} &= \mathbf{F}\big(\omega_{k-1}^{(m)}\big)\,{\boldsymbol{\xi}}_{k-1}^{(m)}, \label{eq:pred_xi}\\
\mathbf{P}_{k|k-1}^{(m)} &= \mathbf{F}\big(\omega_{k-1}^{(m)}\big)\,\mathbf{P}_{k-1}^{(m)}\,\mathbf{F}\big(\omega_{k-1}^{(m)}\big)^T + \mathbf{R}_\xi, \label{eq:pred_P}
\end{align}
with $\mathbf{R}_\xi=\mathrm{diag}(\sigma_c^2\mathbf{I}_2,\sigma_a^2,\sigma_b^2,\sigma_v^2\mathbf{I}_2)$ and
\begin{equation}
\mathbf{F}(\omega) =
\begin{bmatrix}
\mathbf{I}_2 + \omega \Delta t\,\mathbf{J}  & \mathbf{0}_2 & \Delta t\,\mathbf{I}_2 \\
\mathbf{0}_2 & \mathbf{I}_2 & \mathbf{0}_2 \\
\mathbf{0}_2 & \mathbf{0}_2 & \mathbf{I}_2
\end{bmatrix},
\qquad
\mathbf{J} =
\begin{bmatrix}
0 & -1\\
1 & 0
\end{bmatrix}.
\label{eq:Fmat}
\end{equation}
The prediction step propagates $\boldsymbol{\eta}_k$ using \eqref{eq:pred_eta}, and $\boldsymbol{\xi}_k$ using the EKF equations \eqref{eq:pred_xi} and \eqref{eq:pred_P}. The noise terms $r^\omega \sim \mathcal{N}(\mathbf{0}, \sigma_\omega^2)$ and $r^\theta \sim \mathcal{N}(\mathbf{0}, \sigma_\theta^2)$ are drawn from their respective Gaussian distributions to account for process noise.

\subsection{Extended Kalman Filter Update}

The $\boldsymbol{\xi}_k$ component is updated for each particle $m$ using the EKF equations.  The Woodbury identity \cite{woodbury_inverting_1950} is leveraged to avoid inverting the full innovation covariance matrix, which can be large when the number of observations $| \mathcal{P}_k |$ is high. The EKF update equations are given by
\begin{align}
\mathbf{A}_k^{(m)} &= \big(\mathbf{P}_{k|k-1}^{(m)}\big)^{-1} + \sigma_n^{-2}\big(\mathbf{H}_k^{(m)}\big)^T\mathbf{H}_k^{(m)}, \label{eq:A}\\
\mathbf{K}_k^{(m)} &= \sigma_n^{-2}\big(\mathbf{A}_k^{(m)}\big)^{-1}\big(\mathbf{H}_k^{(m)}\big)^T, \label{eq:K}\\
{\boldsymbol{\xi}}_k^{(m)} &= {\boldsymbol{\xi}}_{k|k-1}^{(m)} + \mathbf{K}_k^{(m)}\mathbf{y}_k^{(m)}, \quad
\mathbf{y}_k^{(m)} = \mathbf{z}_k - \mathbf{h}\big({\boldsymbol{\xi}}_{k|k-1}^{(m)},\boldsymbol{\eta}_k^{(m)}\big), \label{eq:xiupd}\\
\mathbf{P}_k^{(m)} &= \big(\mathbf{I}-\mathbf{K}_k^{(m)}\mathbf{H}_k^{(m)}\big)\mathbf{P}_{k|k-1}^{(m)}, 
\label{eq:Pupd}
\end{align}
where $\mathbf{H}_k^{(m)}$ is the Jacobian of the observation function $\mathbf{h}(\cdot)$ with respect to $\boldsymbol{\xi}_k$ evaluated at the predicted state ${\boldsymbol{\xi}}_{k|k-1}^{(m)}$ and the non-linear component $\boldsymbol{\eta}_k^{(m)}$, i.e. 
\begin{equation}
\mathbf{H}_k^{(m)} = \frac{\partial \mathbf{h}}{\partial \boldsymbol{\xi}_k}\big|_{\boldsymbol{\xi}_k = {\boldsymbol{\xi}}_{k|k-1}^{(m)}, \boldsymbol{\eta}_k = \boldsymbol{\eta}_k^{(m)}}. 
\label{eq:H_definition}
\end{equation}

Following the relinearisation principle of the IEKF \cite{liang-qun_iterated_2005} methodology, a single relinearisation refinement step is then performed after the EKF update: $\boldsymbol{h}(\cdot)$ and $\mathbf{H}_k^{(m)}$ are re-evaluated at the updated state ${\boldsymbol{\xi}}_k^{(m)}$ which includes recomputing the specular points in \eqref{eq:specular_point}. 
The resulting relinearised observation model is then used for the particle-weight update described below. 
This relinearisation step also differs from the update procedure used in conventional EKF- and RBPF-based tracking algorithms.

\subsection{Jacobian of the Doppler frequency}

To avoid using numerical differentiation, we derive the Jacobian of the bistatic Doppler frequency \eqref{eq:bistatic_doppler} with respect to the conditionally linear component $\boldsymbol{\xi}_k$ in closed form. 

For the translation speed $\mathbf{v}_k$, the Jacobian is straightforward:
\begin{equation}
\big[\mathbf{H}_k^{(m)}\big]_{(ij), \mathbf{v}_k} = \frac{1}{\lambda}\big( \mathbf{u}_{i,j,k}^{\text{tx}} + \mathbf{u}_{i,j,k}^{\text{rx}} \big)^T, 
\label{eq:H_v}
\end{equation}
with the specular point $\mathbf{s}_{i,j,k}^{(m)}$ in $\mathbf{u}_{i,j,k}^{\text{tx}}$ and $\mathbf{u}_{i,j,k}^{\text{rx}}$ evaluated at the corresponding state ${\boldsymbol{\xi}}_{k}^{(m)}$ and the non-linear component~$\boldsymbol{\eta}_k^{(m)}$.

The dependence of the bistatic Doppler frequency and the other parameters $\{ \mathbf{c}_k, a, b\}$ is via the specular point $\mathbf{s}_{i,j,k}^{(m)}$, which is defined as the stationary solution of the optimisation problem~\eqref{eq:specular_point} along the target boundary. The Jacobian of the specular point with respect to the target state vector $\boldsymbol{\xi}_k$ can thus be derived by implicit differentiation \cite{rudin_principles_1976} of this stationarity condition as 
\begin{equation}
\resizebox{\columnwidth}{!}{$
\big[\mathbf{H}_k^{(m)}\big]_{(ij),\nu}
=
\left.
\nabla_{\mathbf{s}} f_{i,j,k}
\cdot
\frac{\partial\mathbf{s}_{i,j,k}}{\partial\nu}
\right|_{\mathbf{s}_{i,j,k}=\mathbf{s}_{i,j,k}^{(m)}},
\quad
\nu\in\{c_x,c_y,a,b\}
$}
\label{eq:H_implicit}
\end{equation}
with 
\begin{align}
\frac{\partial\mathbf{s}_{i,j,k}}{\partial\nu}
&= -\big(\nabla_{\mathbf{s}}^2 g(\mathbf{s} ; \mathbf{t}_i, \mathbf{r}_j)\big)^{-1}
\cdot \frac{\partial}{\partial\nu}\nabla_{\mathbf{s}} g(\mathbf{s} ; \mathbf{t}_i, \mathbf{r}_j) \\
\nabla_{\mathbf{s}} f_{i,j,k} &= \frac{1}{\lambda}\bigg[
\omega\mathbf{J}^T \big(\mathbf{u}_{i,j,k}^{\text{tx}} + \mathbf{u}_{i,j,k}^{\text{rx}}\big) \nonumber \\
&\quad - \left(\frac{\mathbf{v}_{i,j,k} - (\mathbf{v}_{i,j,k} \cdot \mathbf{u}_{i,j,k}^{\text{tx}})\mathbf{u}_{i,j,k}^{\text{tx}}}{d_{i,j,k}^{\text{tx}}}\right) \nonumber \\
&\quad - \left(\frac{\mathbf{v}_{i,j,k} - (\mathbf{v}_{i,j,k} \cdot \mathbf{u}_{i,j,k}^{\text{rx}})\mathbf{u}_{i,j,k}^{\text{rx}}}{d_{i,j,k}^{\text{rx}}}\right)
\bigg]^T, 
\label{eq:grad_f}
\end{align}
where $\mathbf{J} = \big[\begin{smallmatrix}0&1\\-1&0\end{smallmatrix}\big]$, $d_{i,j,k}^{\text{tx}} = \|\mathbf{s}_{i,j,k}-\mathbf{t}_i\|$ and $d_{i,j,k}^{\text{rx}} = \|\mathbf{s}_{i,j,k}-\mathbf{r}_j\|$ are the distances from the specular point to the transmit and receive antennas, respectively, and $\mathbf{v}_{i,j,k}$ is the velocity of the specular point $\mathbf{s}_{i,j,k}$ as defined in \eqref{eq:speed_specular_point}.

\subsection{Weight Update, Estimation and Resampling}

Using the same Woodbury identity trick as in \eqref{eq:A}-\eqref{eq:Pupd}, the log-likelihood-based weight update for each particle $m$ is computed without inverting the $|\mathcal{P}_k|\times|\mathcal{P}_k|$ innovation covariance as 
\begin{equation}
\begin{aligned}
\log w_k^{(m)}
\propto\;&
\log w_{k-1}^{(m)}
-\frac{1}{2}\Bigl[
\sigma_n^{-2}
\mathbf{y}^{(m)T}\mathbf{y}^{(m)}
\\
&\quad
-\sigma_n^{-4}
\mathbf{y}^{(m)T}
\mathbf{H}_k^{(m)}
\bigl(\mathbf{A}_k^{(m)}\bigr)^{-1}
\mathbf{H}_k^{(m)T}
\mathbf{y}^{(m)}
\Bigr].
\end{aligned}
\label{eq:weight}
\end{equation}
with the innovation $\mathbf{y}_k^{(m)}$ (defined in \eqref{eq:xiupd}) and the Jacobian $\mathbf{H}_k^{(m)}$ (defined in \eqref{eq:H_definition}) evaluated at the updated state ${\boldsymbol{\xi}}_k^{(m)}$ (after relinearisation) and the non-linear component $\boldsymbol{\eta}_k^{(m)}$. 

The weights are then normalised to sum to one, i.e. $w_k^{(m)} \leftarrow w_k^{(m)}/\sum_{m=1}^{N_P} w_k^{(m)}$ and the target state estimate is computed as the weighted mean of the particles, i.e.
\begin{equation}
\hat{\mathbf{x}}_k = \sum_{m=1}^{N_P} w_k^{(m)} \begin{bmatrix} {\boldsymbol{\xi}}_k^{(m)} \\ \boldsymbol{\eta}_k^{(m)} \end{bmatrix}.
\label{eq:state_estimate}
\end{equation}

Systematic resampling \cite{kitagawa_monte_1996} is performed when the effective number of particles $N_\text{eff} = 1/\sum_{m=1}^{N_P} (w_k^{(m)})^2$ falls below a threshold $N_\text{th} = N_P/2$, controlling the particle degeneracy problem. After resampling, the weights are reset to $w_k^{(m)} = 1/N_P$ for all particles.

\cref{alg:rbpf_step} summarises the modified RBPF recursion at time step~$k$.


\begin{table}[t]
   \centering
   \resizebox{\columnwidth}{!}{
   \begin{tabular}{lcc}
   \hline
   & \textbf{Scenario 1} & \textbf{Scenario 2} \\
   \hline
   \textbf{Target trajectory} & Circular & Curved \\
   \textbf{Initial target position} $\mathbf{c}_0$ & $(-25;0)\lambda$ & $(125;125)\lambda$ \\
   \textbf{Initial target velocity} $\mathbf{v}_0$ & $(0;0)\lambda$ & $(-25;55)\lambda$ \\
   \textbf{Initial covariance} $\mathbf{P}_{0}$ & {$\text{diag}(10\lambda,\ 1 a,\ 1 b,\ 5 \lambda,\ 5 |\omega|,\ 5 |\omega| \Delta_t)$} & $0.1 \mathbf{I}_2$ \\
   \textbf{Array} & Circular & Linear \\
   \textbf{Antenna positions} & $\mathbf{t}_i = 50\lambda[\cos(2\pi i/N_T), \sin(2\pi i/N_T)]^T$ & $\mathbf{t}_i = (0; 10 i)\lambda$ \\
   & $\mathbf{r}_j = 50\lambda[\cos(2\pi j/N_R), \sin(2\pi j/N_R)]^T$ & $\mathbf{r}_j = ( 10 j ; 0)\lambda$ \\
   & \multicolumn{2}{c}{$i = 0, \dots, N_T-1$; $j = 0, \dots, N_R-1$} \\
   \hline
   \multicolumn{3}{c}{\textbf{Common parameters}} \\
   \hline
   $a, b$ & \multicolumn{2}{c}{$10\lambda,\ 20\lambda$} \\
   $\omega$ & \multicolumn{2}{c}{$10^\circ/\text{s}$} \\
   $\Delta_t$ & \multicolumn{2}{c}{$0.1$ s} \\
   $N_t, N_r$ & \multicolumn{2}{c}{50, 50} \\
   Mean SNR & \multicolumn{2}{c}{20 dB} \\
   $\mathbf{Q}$ & \multicolumn{2}{c}{$\text{diag}(0.01\lambda,\ 0.01 a,\ 0.01 b,\ 0.01 \lambda,\ 0.01 |\omega|,\ 0.01 |\omega| \Delta_t)$} \\
   \textbf{Particles State Initialisation} & \multicolumn{2}{c}{$\mathbf{x}_0^{(m)} \sim \mathcal{N}(\mathbf{x}_0, \mathbf{P}_{0})$} \\
   \hline
   \end{tabular}
   }
   \caption{Simulation parameters for different scenarios.}
   \label{tab:sim_params}
\end{table}

\section{Numerical Results}

The performance of the proposed modified RBPF algorithm is evaluated through numerical simulations. Two scenarios are considered, with different target trajectories and antenna configurations, as summarised in \cref{tab:sim_params} (unless stated otherwise). The results are averaged over $100$ Monte Carlo runs. 




\begin{figure}
\centering
\includegraphics[width=0.65\linewidth, trim=55 30 50 35, clip]{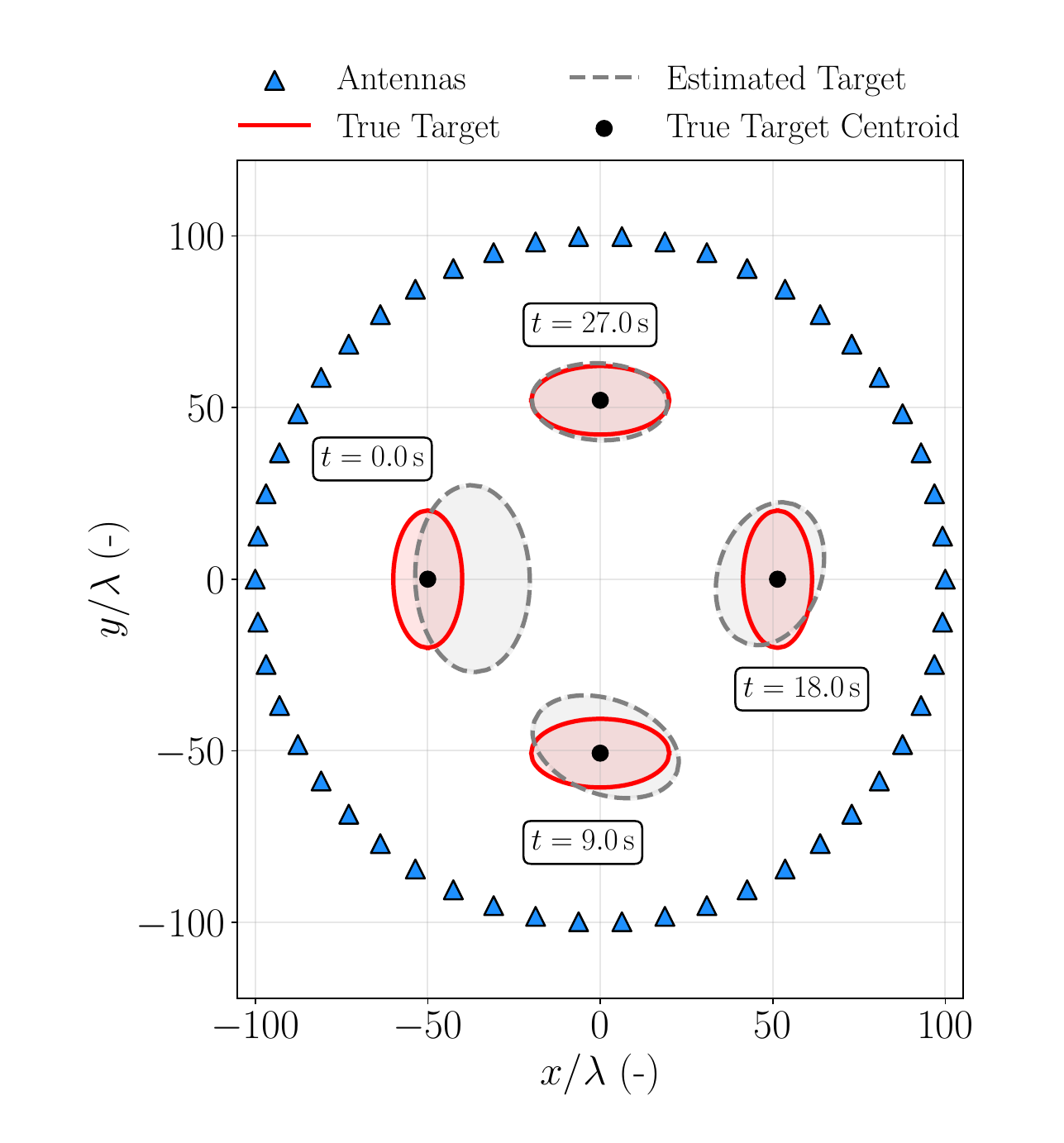}
\caption{Tracking process over time for the extended target model. Scenario 1 is considered, with $N_P=40$ particles.}
\label{fig:qualitative_representation_time}
\end{figure}


Fig.~\ref{fig:qualitative_representation_time} illustrates the tracking process over time for the extended target model of Scenario 1 in \cref{tab:sim_params}. The true target is represented by a red ellipse, while the estimated target is shown as a grey ellipse. As time passes, the extended target model accurately tracks the target position and orientation. 
The figure highlights the effectiveness of the extended target model in capturing the target state, which is crucial for accurate tracking in scenarios where the target cannot be approximated as a point.

In the following sections, we further evaluate the performance of the proposed modified RBPF algorithm in terms of tracking accuracy and robustness, considering different numbers of particles and comparing it with other tracking algorithms.



\subsection{Impact of the Number of Particles}
\label{sec:impact_of_the_number_of_particules}


\begin{figure}
\centering
\includegraphics[width=\linewidth, trim=25 30 25 0, clip]{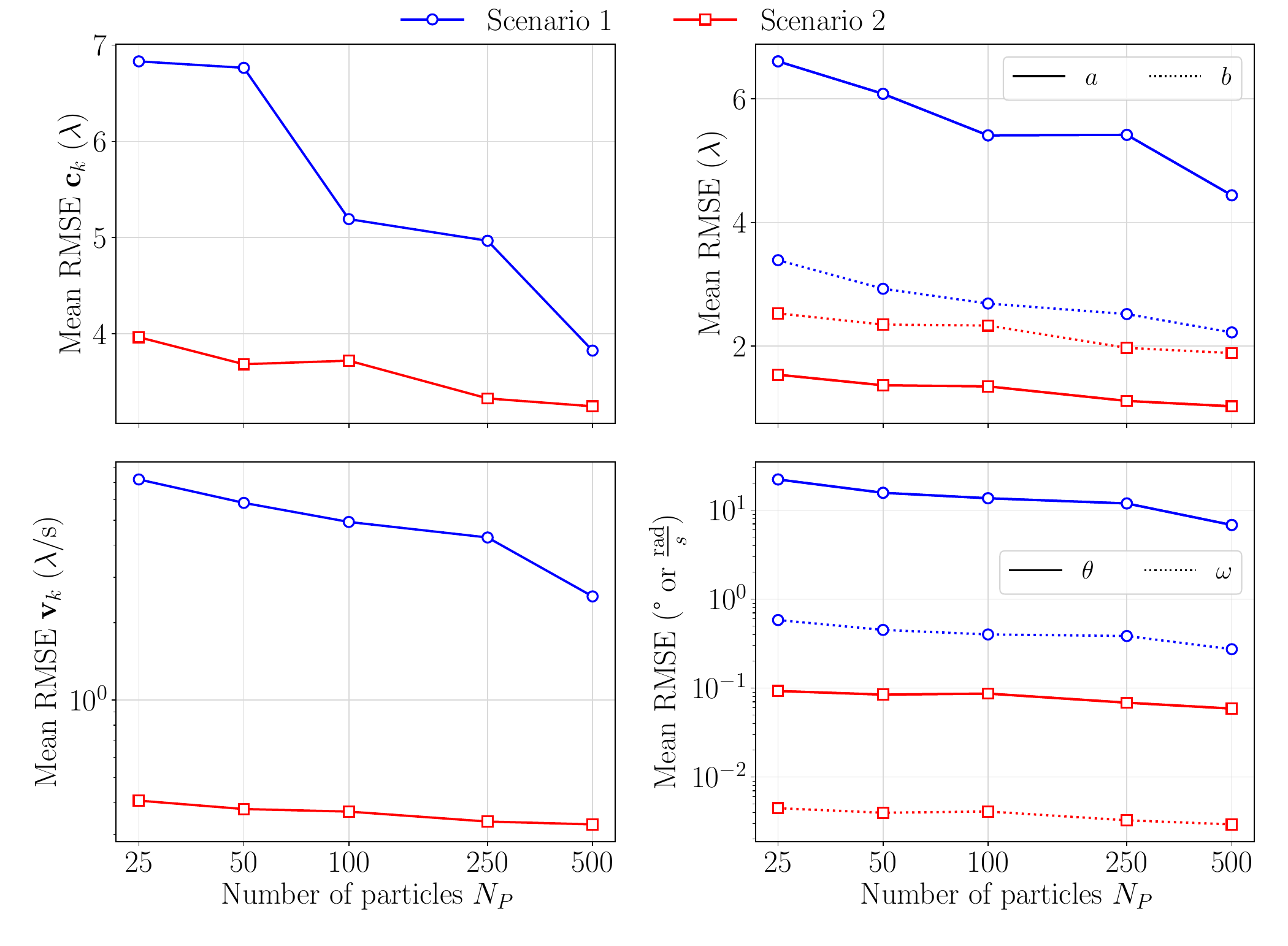}
\caption{RMSE of the state parameters $\mathbf{x}_k$ as a function of the number of particles $N_P$ for both scenarios. }
\label{fig:rmse_vs_n_particules}
\end{figure}

Fig.~\ref{fig:rmse_vs_n_particules} shows the mean RMSE of the target state parameters $\mathbf{x}_k$ versus the number of particles $N_P$ for both scenarios. The tracking performance improves (lower RMSE) as the number of particles increases (as expected since more particles provide a better approximation of the posterior distribution of the target state) but only slightly from $N_P=25$ to $N_P=500$. This indicates that only a limited number of particles is required to approach the tracking performance obtained with much larger particle sets, favorable for real-time applications given the linear complexity in $N_P$ of the proposed RBPF algorithm.

\subsection{Tracking Algorithms Comparison}
\label{sec:tracking_algorithm_comparison}

\begin{figure}
   \centering

   \begin{subfigure}{1\linewidth}
      \centering
      \includegraphics[width=\linewidth, trim=25 30 30 55, clip]{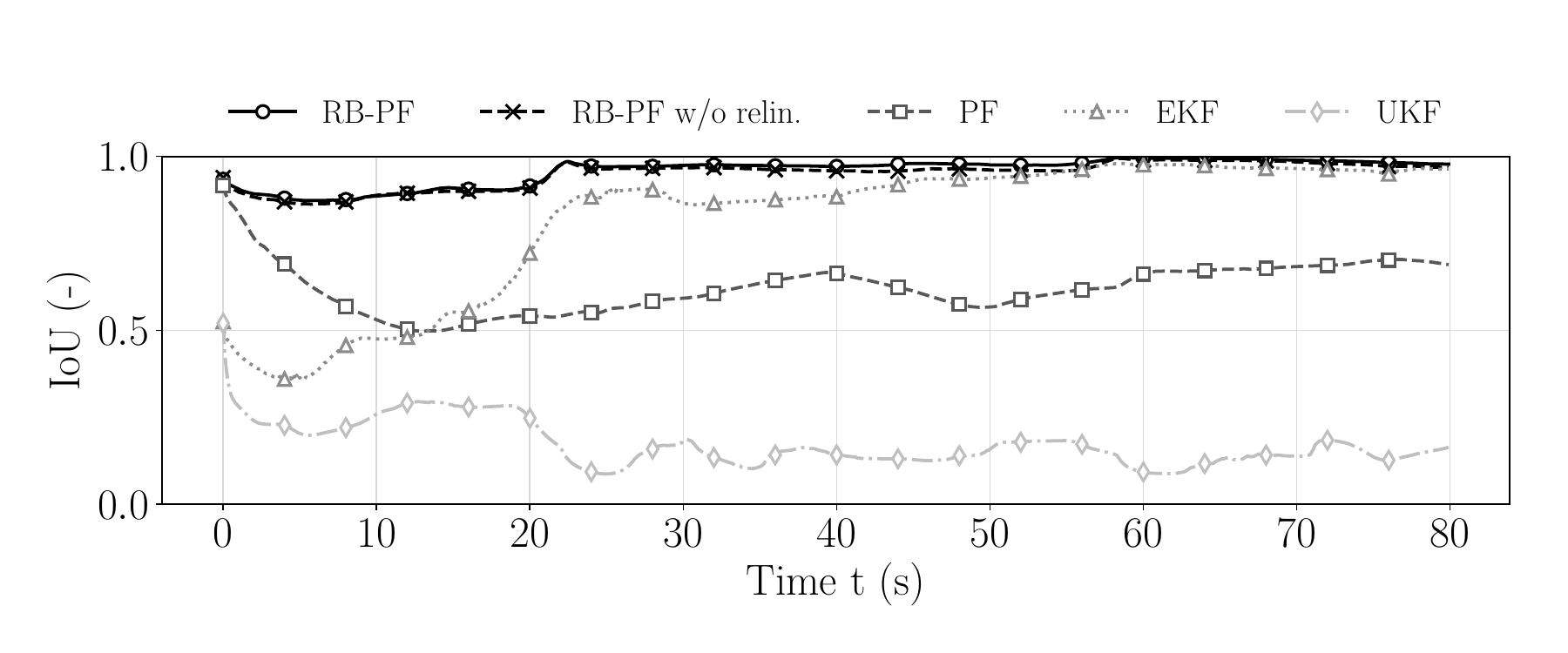}
   \caption{Scenario 1.} 
   \label{}
   \end{subfigure}

   \begin{subfigure}{1\linewidth}
      \centering
      \includegraphics[width=\linewidth, trim=25 30 30 70, clip]{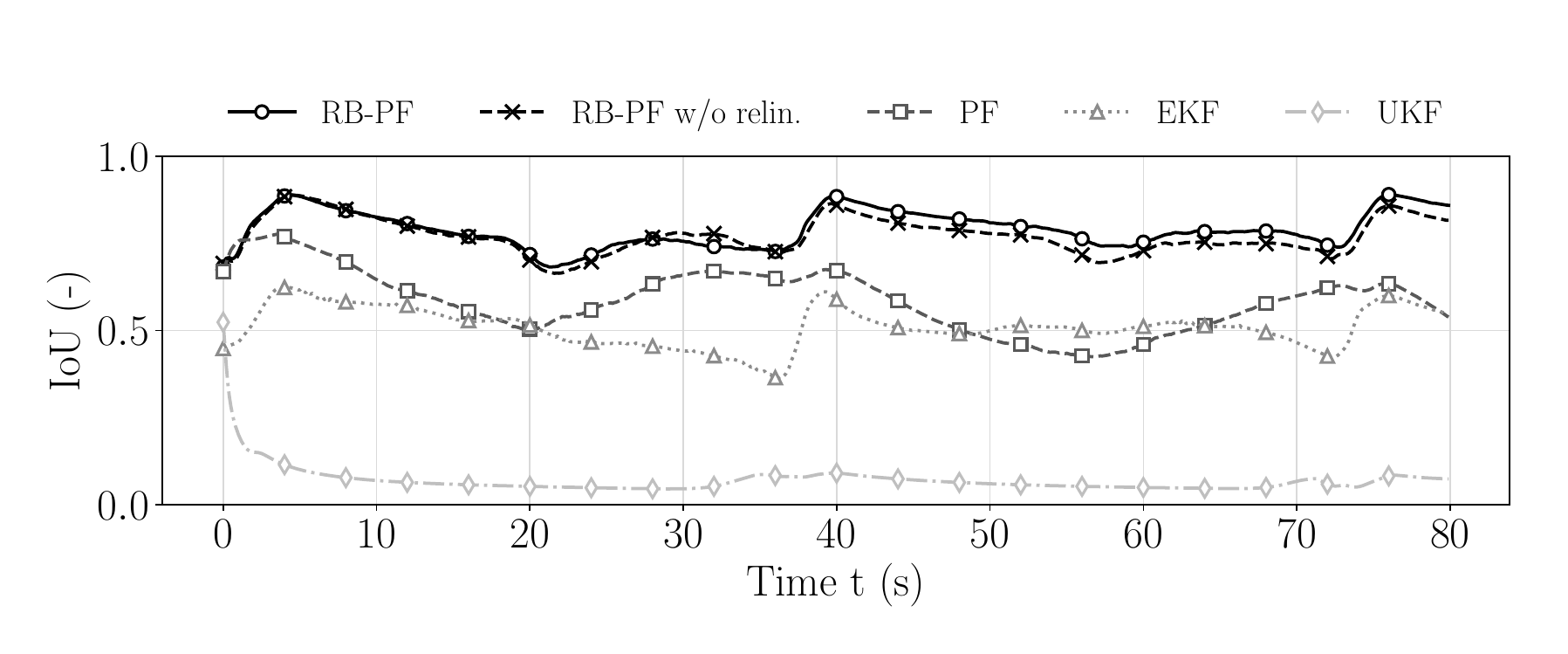}
   \caption{Scenario 2.} 
   \label{}
   \end{subfigure}

\caption{Comparison of the Intersection over Union (IoU) for the proposed RBPF and other tracking algorithms in both scenarios. }
\label{fig:tracking_algorithm_comparison}
\end{figure}

\begin{figure}
\centering
\includegraphics[width=1\linewidth, trim=20 30 25 55, clip]{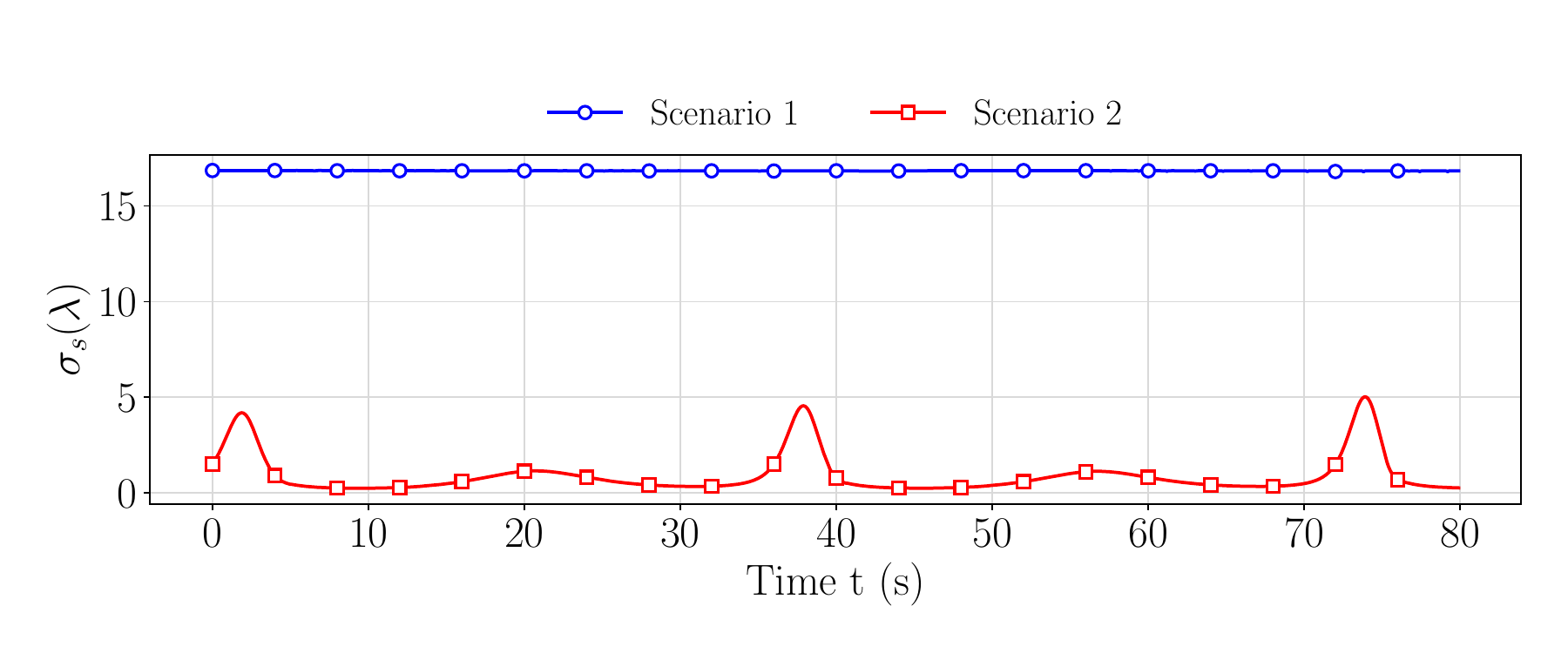}
\caption{Standard deviation $\sigma_{s}$ 
of the specular points $\mathbf{s}_{i,j}$ for both scenarios along the target trajectory. }
\label{fig:variance_points_speculaires}
\end{figure}

\begin{table}[t]
\centering
\scriptsize
\setlength{\tabcolsep}{2pt}
\renewcommand{\arraystretch}{1.25}

\begin{tabularx}{\columnwidth}{|>{\centering\arraybackslash}p{0.12\columnwidth}|
                                  >{\centering\arraybackslash}X|
                                  >{\centering\arraybackslash}X|
                                  >{\centering\arraybackslash}X|}
\hline
\textbf{Algorithm} &
\textbf{Specular Points} &
\textbf{Update} &
\textbf{Dominant Complexity} \\
\hline

\textbf{RBPF} &
$\mathcal{O}(N_PN_aN_{\mathrm{scan}})$ &
$\begin{gathered}
\mathcal{O}(N_PN_an_{\boldsymbol{\xi}}\\
{}+N_P(n_{\boldsymbol{\xi}}^3+
N_an_{\boldsymbol{\xi}}^2))
\end{gathered}$ &
$\mathcal{O}(N_PN_aN_{\mathrm{scan}})$ \\[3mm]

\hline

\textbf{PF} &
$\mathcal{O}(N_PN_aN_{\mathrm{scan}})$ &
$\mathcal{O}(N_PN_a)$ &
$\mathcal{O}(N_PN_aN_{\mathrm{scan}})$ \\[2mm]

\hline

\textbf{EKF} &
$\mathcal{O}(N_aN_{\mathrm{scan}})$ &
$\mathcal{O}(N_an+n^3+N_an^2)$ &
$\mathcal{O}(N_aN_{\mathrm{scan}})$ \\[2mm]

\hline

\textbf{UKF} &
$\mathcal{O}((2n+1)N_aN_{\mathrm{scan}})$ &
$\mathcal{O}((2n+1)^3+(2n+1)N_a)$ &
$\mathcal{O}((2n+1)N_aN_{\mathrm{scan}})$ \\[2mm]

\hline
\end{tabularx}

\caption{Computational complexity of the considered tracking algorithms
for a single time step $k$. $N_a=N_TN_R$, $N_{\mathrm{scan}}$ is the
number of iterations for the specular-point optimisation, $n=8$ is the
state dimension, and $n_{\boldsymbol{\xi}}=6$ is the dimension of the
conditionally linear state.}
\label{tab:algo_complexity}
\end{table}

Fig.~\ref{fig:tracking_algorithm_comparison} compares the performance of the proposed modified RBPF against standard Particle Filter (PF) \cite{doucet_sequential_2013}, Extended Kalman Filter (EKF) \cite{snyder_introduction_2014} and Unscented Kalman Filter (UKF) \cite{julier_unscented_2004} (all adapted to our extended target tracking scenario) on both scenarios. Their computational complexity is compared in \cref{tab:algo_complexity}. The Intersection over Union (IoU) metric is used to evaluate the tracking performance, which measures the overlap between the estimated target shape and the ground truth compared to their union. For fair comparison between the two trajectories, both scenarios use the same covariance $\mathbf{P}_{0}$ of Scenario 2. To isolate the benefit of Rao-Blackwellisation itself, the PF baseline is allocated the same number of particles $N_P$ as the proposed RBPF. 

Overall, lower IoU values are achieved in Scenario 2 compared to Scenario 1. As shown in Fig.~\ref{fig:variance_points_speculaires}, the standard deviation of the specular points $\mathbf{s}_{i,j}$ along the target trajectory is lower in Scenario 2, indicating denser and less diverse observations, which makes the tracking problem more challenging.

Due to the more complex target trajectory of Scenario 2, the EKF performance, close to that of the RBPF in Scenario~1, is significantly degraded while the PF performance is less affected by the target trajectory complexity, as it can better handle higher non-linearities in the state-space model. 
The proposed modified RBPF, combining PF propagation with EKF updates, outperforms the other tracking algorithms in both scenarios, achieving higher IoU values along the target trajectories.

The IoU improvement brought by the relinearisation step (step 7 in \cref{alg:rbpf_step}) can also be observed in Fig.~\ref{fig:tracking_algorithm_comparison}, where the RBPF with relinearisation outperforms the one without, particularly in Scenario 2 with the curved target trajectory. 
This highlights the benefit of relinearisation, which enables more accurate weight updates by re-evaluating the observation model at the updated conditionally linear state $\boldsymbol{\xi}_k$.

Table~\ref{tab:algo_complexity} summarises the computational complexity of the four tracking algorithms for a single time step $k$. Interestingly, the modified RBPF shares the same dominant complexity order as the standard PF, $\mathcal{O}(N_P N_a N_\text{scan})$, since both are governed by the specular point search repeated for each of the $N_P$ particles. The additional EKF-related terms in the RBPF remain lower order in the considered scenarios. The accuracy improvement of the RBPF over the PF observed in Fig.~\ref{fig:tracking_algorithm_comparison} is thus obtained at a relatively low additional computational cost. 
Conversely, the EKF and UKF have a lower, particle-independent complexity, but suffer from degraded tracking accuracy, particularly in Scenario 2, where the strong non-linearities induced by the curved trajectory are poorly captured by a single Gaussian approximation. Together with the results of \cref{sec:impact_of_the_number_of_particules}, showing that a moderate particle number ($N_P \approx 25\text{--}50$) is sufficient for good tracking performance, this highlights the advantage of the proposed RBPF algorithm in terms of accuracy-complexity trade-off. 

Comparing the IoU evolution of the RBPF algorithm with the specular point standard deviation in Fig.~\ref{fig:variance_points_speculaires} also suggests a correlation between tracking performance and observation diversity. For instance, in Scenario 2, IoU increases following rises in specular point dispersion (e.g., after $\text{t}={2,38,74}$ s).

\subsection{Impact of the SNR}
\label{sec:impact_of_the_snr}

\begin{figure}
\centering
\includegraphics[width=0.9\linewidth, trim=0 30 0 70, clip]{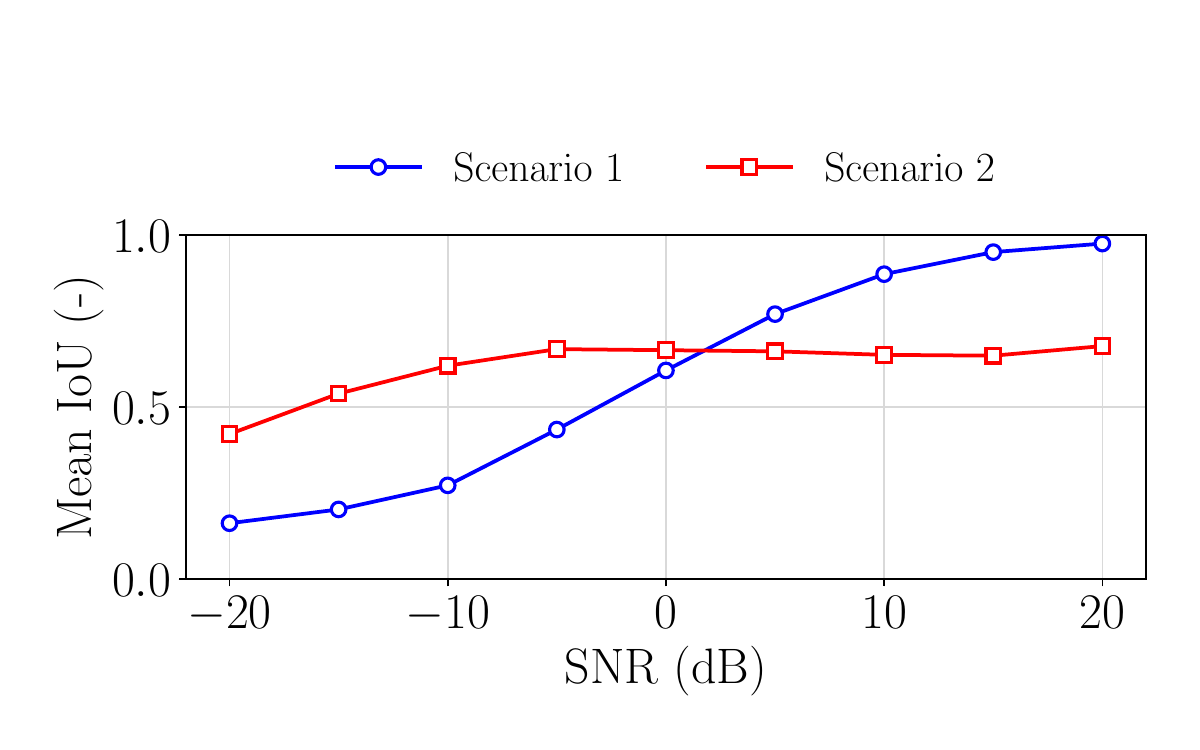}
\caption{Mean IoU metric as a function of the mean SNR for both scenarios.}
\label{fig:mean_iou_vs_snr}
\end{figure}

Fig.~\ref{fig:mean_iou_vs_snr} shows the mean IoU as a function of the mean SNR for both scenarios and reveals two distinct operating regimes. For Scenario 1, where the specular points are more diverse (see Fig.~\ref{fig:variance_points_speculaires}), the IoU is monotonically increasing with the SNR, approaching $\text{IoU} \approx 1$ at high SNR, while it is quite low for low SNR values, indicating that performance is mainly limited by measurement noise. 
In contrast, for Scenario 2, the IoU quickly saturates around $\text{IoU} \approx 0.65$ beyond $\text{SNR} = 5 \text{dB}$, confirming that performance is limited by the lack of geometric diversity of the specular points, as shown in Fig.~\ref{fig:variance_points_speculaires}. In this case, increasing the SNR provides little improvement, as the observations remain insufficiently informative for accurate state estimation.

The two curves intersect around $\text{SNR}\approx0$--$2$ dB, marking the transition between the two regimes: Scenario 2 is more robust at low SNR thanks to its denser observation set, whereas Scenario 1 achieves better performance once measurement noise is no longer dominant since it is not capped by any geometric ceiling. 
This distinction matters for system design, as it indicates that improving the SNR may not always yield better tracking performance if the observation geometry is not sufficiently diverse.


\section{Conclusion}

This paper developed a novel Rao-Blackwellised particle filter for tracking extended targets using solely bistatic Doppler frequency measurements, combining particle filtering for the non-linear state components with particle-specific Extended Kalman filtering for the other ones, along with a post-update relinearisation step to improve the particle-weight evaluation.
Simulations show that the proposed algorithm better handles the nonlinearities of the state-space model, improving tracking performance over standard particle and Kalman-based methods at a relatively low additional computational cost.

Future work will investigate the impact of the antenna array geometry on tracking performance and extend the proposed framework to multiple extended targets.


\bibliographystyle{ieeetr}
\bibliography{biblio}

\end{document}